\documentclass[twocolumn,showpacs,preprintnumbers,amsmath,amssymb,floatfix,a4paper,fleqn,prl]{revtex4-1}

\usepackage[dvips]{graphicx}

\usepackage{dcolumn}
\usepackage{bm}
\usepackage{amsmath}
\usepackage{amssymb}
\usepackage{color}
\usepackage{units}

\usepackage[french]{babel}
\usepackage[latin1]{inputenc}

\long\def\symbolfootnote[#1]#2{\begingroup \def\thefootnote{\fnsymbol{footnote}}\footnote[#1]{#2}\endgroup}

\begin{document}
\preprint{APS/}

\title{Stability of ion confinement for a novel mass spectrometer of infinite mass range}

\author{Alexandre Vallette}\email{alexandre.vallette@spectro.jussieu.fr}
\author{C. I. Szabo} \email{csilla.szabo@spectro.jussieu.fr }
\author{P. Indelicato} \email{paul.indelicato@lkb.ens.fr}

\affiliation{
Laboratoire Kastler Brossel, École Normale Supérieure,
CNRS, Université Pierre et Marie Curie -- Paris 6, Case 74; 4, place
Jussieu, 75252 Paris CEDEX 05, France}

\date{October 24, 2012}

\begin{abstract}

We study the ions dynamics inside an Electrostatic Ion Beam Trap (EIBT) and show that the stability of the trapping is ruled by a Hill's equation. This unexpectedly demonstrates that an EIBT, in the reference frame of the ions works very similar to a quadrupole trap. The parallelism between these two kinds of traps is illustrated by comparing experimental and theoretical stability diagrams of the EIBT. The main difference with quadrupole traps is that the stability depends only on the ratio of the acceleration and trapping electrostatic potentials, not on the mass nor the charge of the ions. All kinds of ions can be trapped simultaneously and since parametric resonances are proportional to the square root of the charge/mass ratio the EIBT can be used as a mass spectrometer of infinite mass range.

\end{abstract}
\pacs{07.75.+h,37.10.Ty,82.80.Ms}
\keywords{electrostatic trap, parametric resonance, stability parameter, hill's equation,mathieu equation, mass spectrometry, EIBT, stability map}
                              
\maketitle

Electrostatic Ion Beam Traps \cite{zhv97} are taking an important place in between very low-energy charged-particles storage devices, such as quadrupole and Penning traps \cite{MGW05} and high energy storage rings \cite{LAR1995}. With the ConeTrap \cite{SCJ2001}, electrostatic rings \cite{mol1997} and the Mini-Ring \cite{bmbt2008}, they form a new family of traps operating at energies of a few keV. They are used for atomic and molecular metastable-states studies, molecular fragmentation and photodissociation (see, e.g., \cite{ahz2004} for a review). Beyond providing trapping of energetic particles in a well defined direction, these traps have many interesting features: they are small, relatively inexpensive, easy to setup and operate and have a field-free region where ions move freely and where measurements can easily be performed. They can even be used as Time Of Flight (TOF) mass spectrometers \cite{sgp2002} or cooled at cryogenic temperatures \cite{LFM2010}.

Despite these interesting features, all the published theoretical models describing EIBT are based on one dimensional approximations, neglecting the radial motion. This leads to inaccurate predictions of the trap stability and operating domain, and restricts their flexibility, as finding reasonable working points requires lengthy and tedious experimental exploration. This may partly be explained by the lack of an analytical formula for the electrostatic potential inside these traps leading to the dilemma of choosing between a simplistic analytical model and a heavy numerical treatment unsuited to explore the huge space of parameters. Usual beam simulation codes fail to produce good results as the numerical inaccuracies at the ion turning points lead to energy non conservation reaching a few  \unit{100} {eV} over  a  few tens of oscillations. Here we solve the problem, using  methods developed for radio-frequency quadrupole traps. We show that the radial dynamic is ruled by a Hill's equation, a particular case of the Mathieu equation that describes quadrupole traps. This model yields accurate predictions of the ions motion in the trap and of the stability region. 

\begin{figure}
\includegraphics[width=0.5\textwidth,trim =5cm 5.8cm 1cm 4.7cm,clip]{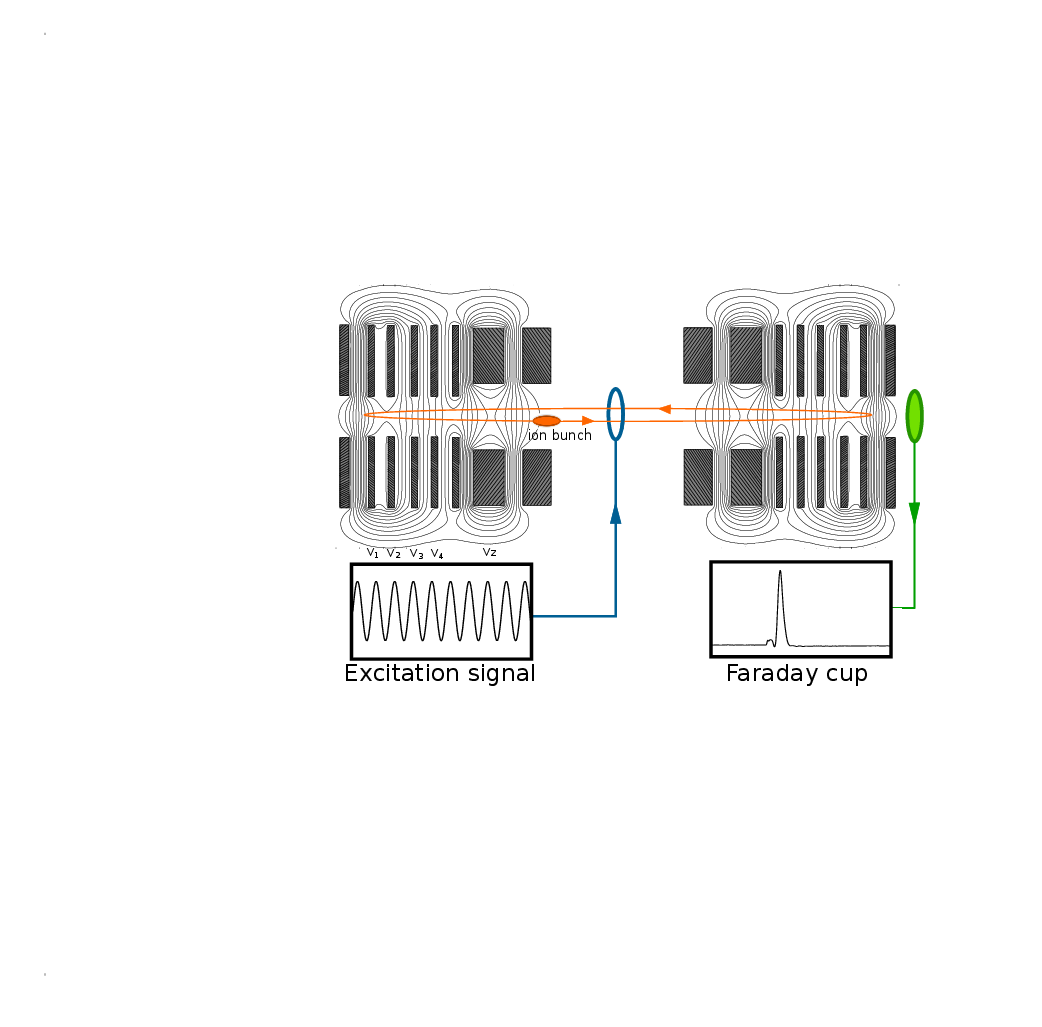}
\caption{\label{setup} (Color online) Overview of the experimental setup. Five potentials are applied to the electrodes (striped rectangles with $V_i$), the others are grounded. The resulting electrostatic field is represented by contour lines. The injection of the ion bunch (orange) is performed when all the electrodes on one side of the trap are grounded. The potentials are raised before the bunch has time to come back so that the ions go back and forth between the two stacks of electrodes. The faraday cup (green) is linked to an oscilloscope via a charge amplifier. A sinusoidal voltage can be applied to a central electrode via a signal generator.}
\end{figure}

The design and operation of the EIBT has been described previously in detail \cite{zhv97} and a schematic drawing of the ion trap is shown in Fig.~\ref{setup}. The trap consists of two sets of coaxial cylindrical electrodes roughly equivalent to two spherical mirrors, the electrostatic analog of a Fabry-Perot interferometer. The configuration of the trap is defined by the potentials of five of these electrodes $\left\{V_1,V_2,V_3,V_4,V_z\right\}$ on both ends of the trap, the others being grounded. The length of the trap is \unit{422} {mm} and the inner radius of the electrodes varies from \unit{8} {mm} to \unit{13} {mm}. An oscillating potential $V_{ex}\sin(\omega_{ex}t)$ can be applied to a hollow electrode located at the center of the trap, in order to destabilize ion trajectories as we will see in the sequel. Both $V_{ex}$ (a few volts) and $\omega_{ex}$ (a few MHz) are adjustable.

The large number of parameters implied in the tuning of the EIBT makes changing of the setup difficult as it requires a lengthy trial and error procedure. In a previous article \cite{VI2010}, we presented a method to obtain an analytical formula of the potential inside the EIBT. Using this method, we calculate the potential on the axis $V(z)$ as a function of  the five potentials applied to the electrodes. The general solution of the Laplace equation in cylindrical symmetry, is a series of the form
\begin{equation}
\label{quadpot}
V(r,z)=\sum_{n=0}^{+\infty}\frac{(-1)^n}{2^{2n}(n!)^2 }r^{2n}V^{(2n)}(z).
\end{equation}
Similar to the method with Paul traps, we limit ourselves to quadratic terms, which decouples the radial and axial motion and gives a linear model. The trajectory of an ion in the trap, without any external excitation, is described by the following set of equations:
\begin{subequations} \label{E:gp}
\begin{gather}
\frac{d^2z}{dt^2}=\left(-\frac{dV(z)}{dz}+\frac{1}{4}r^2 \frac{d^3V(z)}{dz^3}\right) \label{E:gp1} \\
\frac{d^2r}{dt^2}=\frac{1}{2}r \frac{d^2V(z)}{dz^2} \label{E:gp2},
\end{gather}
\end{subequations}
where, in order to have dimensionless equations, we made the following substitutions: $z \to \frac{z}{L}$, $r \to \frac{r}{L}$, $t \to \frac{t}{\tau}$  and $V_i \to \frac{V_i}{E}$, where $L$ is the half-length of the trap, $v_0=\sqrt{\frac{2qE}{m}}$, $E$ is the acceleration voltage and $\tau=\frac{L}{v_0}$. 

With $z(0)=0$, $r(t)=0$, $\frac{dz}{dt}(t=0)=1$, Eq.~\eqref{E:gp1} can be solved and describes the motion $z(t)$ along the $z$-axis, which is periodic of period $T=1/f_z$. Substituting in Eq.~\eqref{E:gp2}, we obtain the following Hill's equation \cite{HAK1991}:
\begin{equation}
\label{hill}
\frac{d^2r}{dt^2}-\left(\frac{1}{2} \frac{d^2V(z(t))}{dz^2}\right) r=0\,.
\end{equation}
This corresponds to a change of the reference frame from the trap's to the ion's. The longitudinal motion of the ion plays here the same role as the quadrupole traps radiofrequency.
The principal matrix of \eqref{hill} is 
\begin{equation}
M(t)=\left( 
\begin{array}{cc}
\psi_1(t,t_0)&\psi_2(t,t_0)\\
\dot{\psi_1}(t,t_0)&\dot{\psi_2}(t,t_0)
\end{array}
\right),
\end{equation}
where $\psi_1(t,t_0)$ is the solution of Eq.~\eqref{hill} with initial conditions $\psi_1(t_0,t_0)=1$ and $\dot{\psi_1}(t_0,t_0)=0$, and $\psi_2(t,t_0)$ with $\psi_2(t_0,t_0)=0$ and $\dot{\psi_2}(t_0,t_0)=1$ respectively. Liouville's formula \cite{HAK1991} shows that $\det M(t,t_0)=1$ and therefore the characteristic equation of the monodromy matrix $M(t_0+T)$ is given by $x^2-2\Delta x+1=0$ where:
\begin{equation*}
\Delta=Tr(M(t_0+T))=\frac{\psi_1(t_0+T,t_0)+\dot{\psi_2}(t_0+T,t_0)}{2}.
\end{equation*}
Applying Floquet's theorem \cite{HAK1991}, we know that if $\Delta^2>1$, one of the two solutions is unbound, while for $\Delta^2<1$ there are two bounded solutions of the form:
\begin{equation}
\label{floquet_sol}
r(t)=e^{\pm i \frac{\pi}{T}\beta t}p_{\pm}(t),
\end{equation}
where $p_{\pm}(t+T)=p_{\pm}(t)$ is periodic of period $T$ and $ \Delta = cos(\pi \beta)$. $\beta$ is often called the stability parameter \cite{MGW05}.
\par To sum up, Hill's equation is stable, and thus trapping can be observed, when $| \Delta |<1$, which is equivalent to $0+2k\pi<\beta<1+2k\pi\quad\forall k \in \mathbb{Z}$.

We performed experiments using  ion beams produced by the SIMPA\cite{gtas2010} \unit{14.5} {GHz} electron cyclotron resonance ion source (ECRIS), accelerated to $E=$ \unit{5.2} {keV/charge}. A magnetic dipole enables us to select the ions by their mass/charge ratio. An ion beam is continuously injected into the trap, in a potential configuration corresponding to no entrance mirror and a closed output mirror. A sequencer is used to activate a set of fast high-voltage switches to close the entrance mirror, trapping the ions, which then oscillate between the two mirrors with a period of $\approx 2\mu$s. The number of ions remaining in the trap after a fixed time (typically \unit{500} {{$\mu$s}}) can be measured by switching off the exit mirror: particles are ejected out of the trap and hit a faraday cup. The signal is processed by a charge pre-amplifier and a double delay line  amplifier. After calibration, the uncertainty on the number of ions is $\approx$10\%. In this article, we present experimental data that was observed with $Ne^{5+}$ at \unit{5.2} {kV/charge}. We varied only two of the five potentials, $V_1$ and $V_z$, the three others, $V_2$, $V_3$ and $V_4$ being fixed at a constant value, respectively \unit{5.85} {kV}, \unit{4.15} {kV} and \unit{1.65} {kV}. 

\begin{figure}
\includegraphics[width=\columnwidth,trim =0.8cm 1.cm 0.1cm 1.cm,clip]{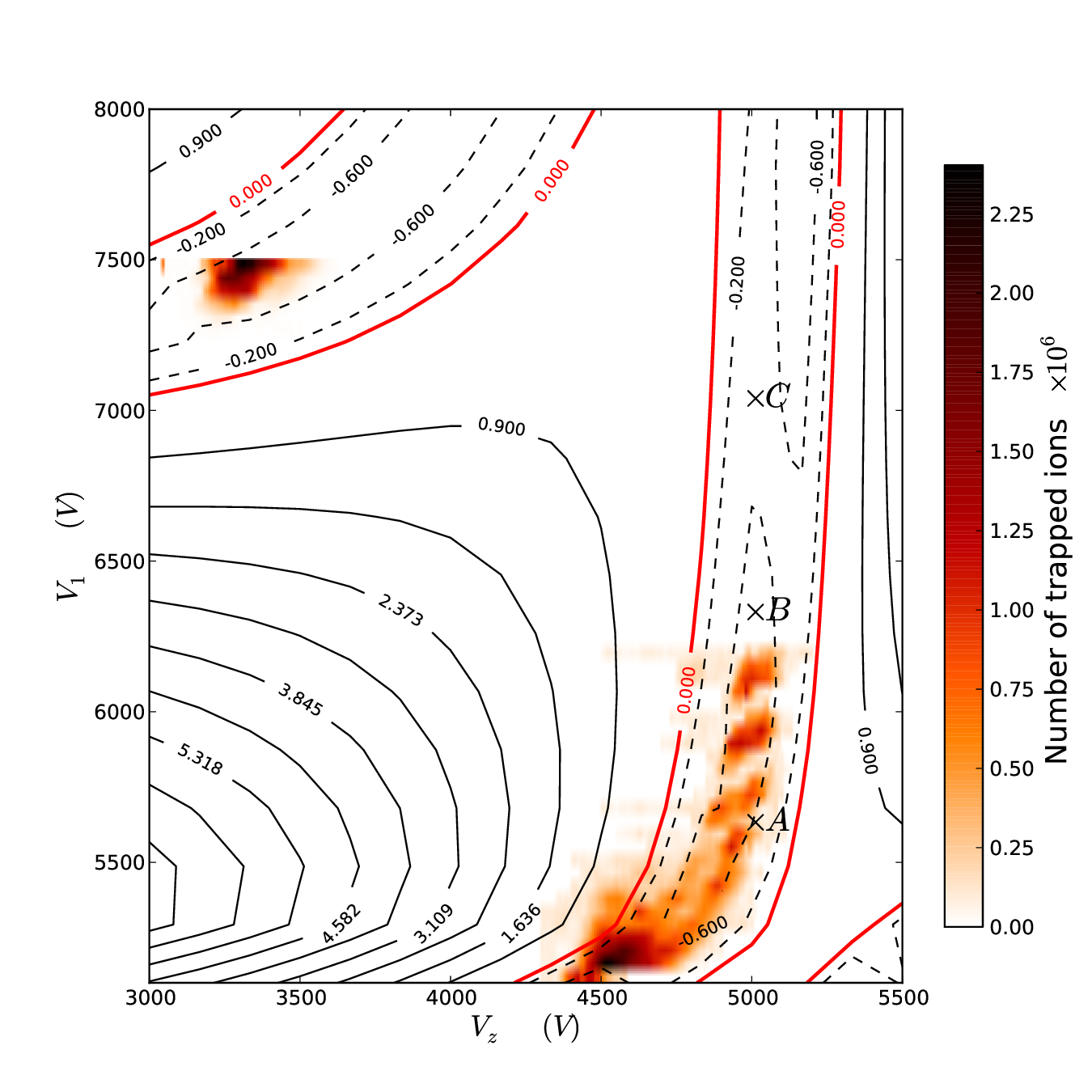}
\includegraphics[width=\columnwidth,trim =1cm 4.3cm 1cm 5cm,clip]{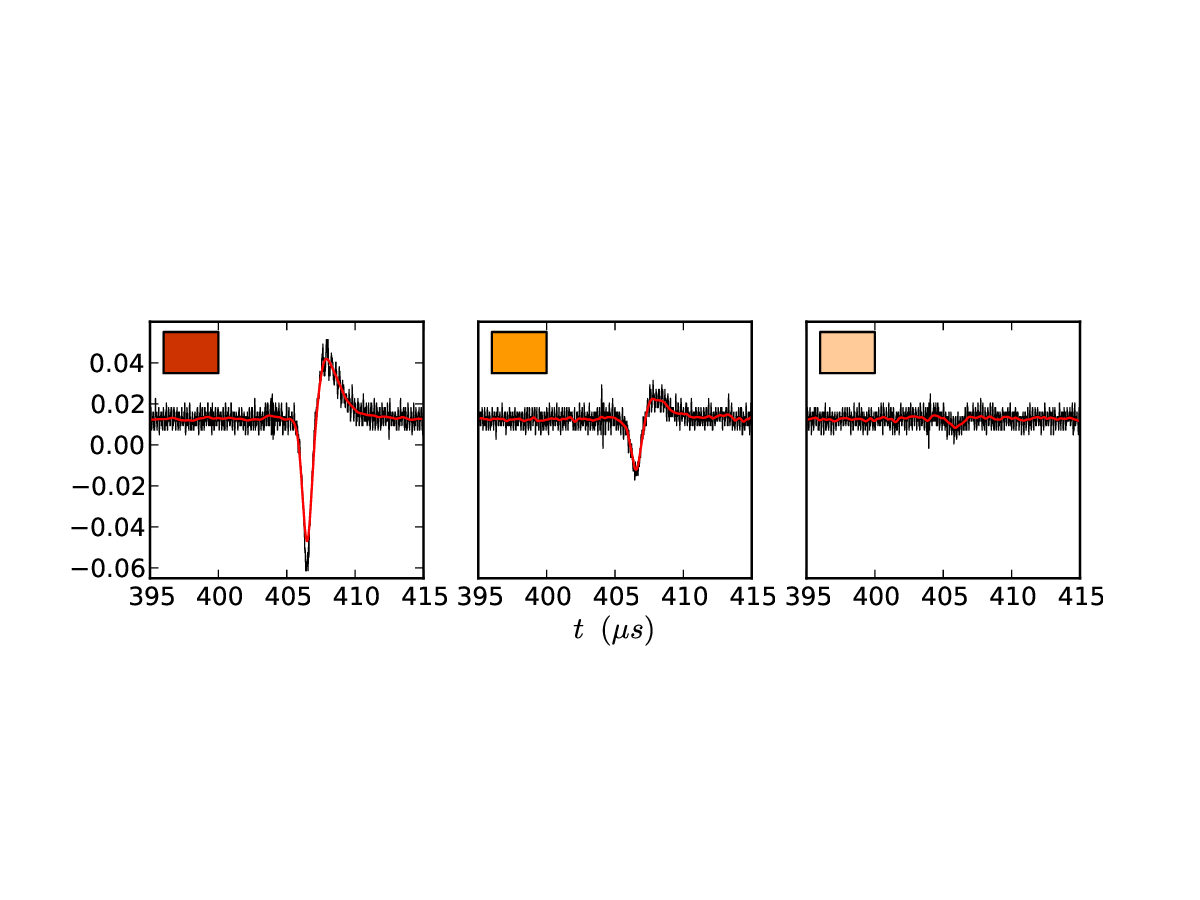}
\caption{\label{stab} (Color online) The upper part is the stability diagram, a comparison between theory and experiment: contour lines represent constant values of the stability coefficient $|\Delta|-1$. The stability regions predicted by theory, defined by $| \Delta |-1<0$, are the two narrow areas between the red lines. The orange shaded map represents the number of ions remaining after \unit{500} $\mu$s of experimental trapping. That is for each point with abscissa $V_z$ and ordinate $V_1$, we trapped in the resulting configuration and the more ions remained, the more colored is the point. The signal recorded from the charge amplifier connected to the faraday cup is presented in the lower part of the figure for three different cases, together with the color code used in the upper part.}
\end{figure}

Figure \ref{stab} shows a comparison between the theory outlined above and experiment. On one hand, the contours computed from theory, show constant values $| \Delta |-1$. The stability regions being defined by $| \Delta |-1<0$, we see that there are two of them lying between the red lines where contours are dashed. On the other hand, the shaded map represents the experimental number of ions remaining after \unit{500} {$\mu$s} of trapping in a given configuration of ($V_z$,$V_1$). For each point, we made several trapping cycles to obtain the average number of ions trapped in this configuration. The resolution of the experimental map is \unit{10} {V} along $V_z$ and  \unit{50} {V} along $V_1$. Because we need to switch high voltages in less than  \unit{1} {{$\mu$s}}, we are are limited to a maximum value of $V_1$ of \unit{7500} {V}.

\begin{figure}
\includegraphics[width=\columnwidth,trim =1.5cm 3.9cm 2cm 5cm,clip]{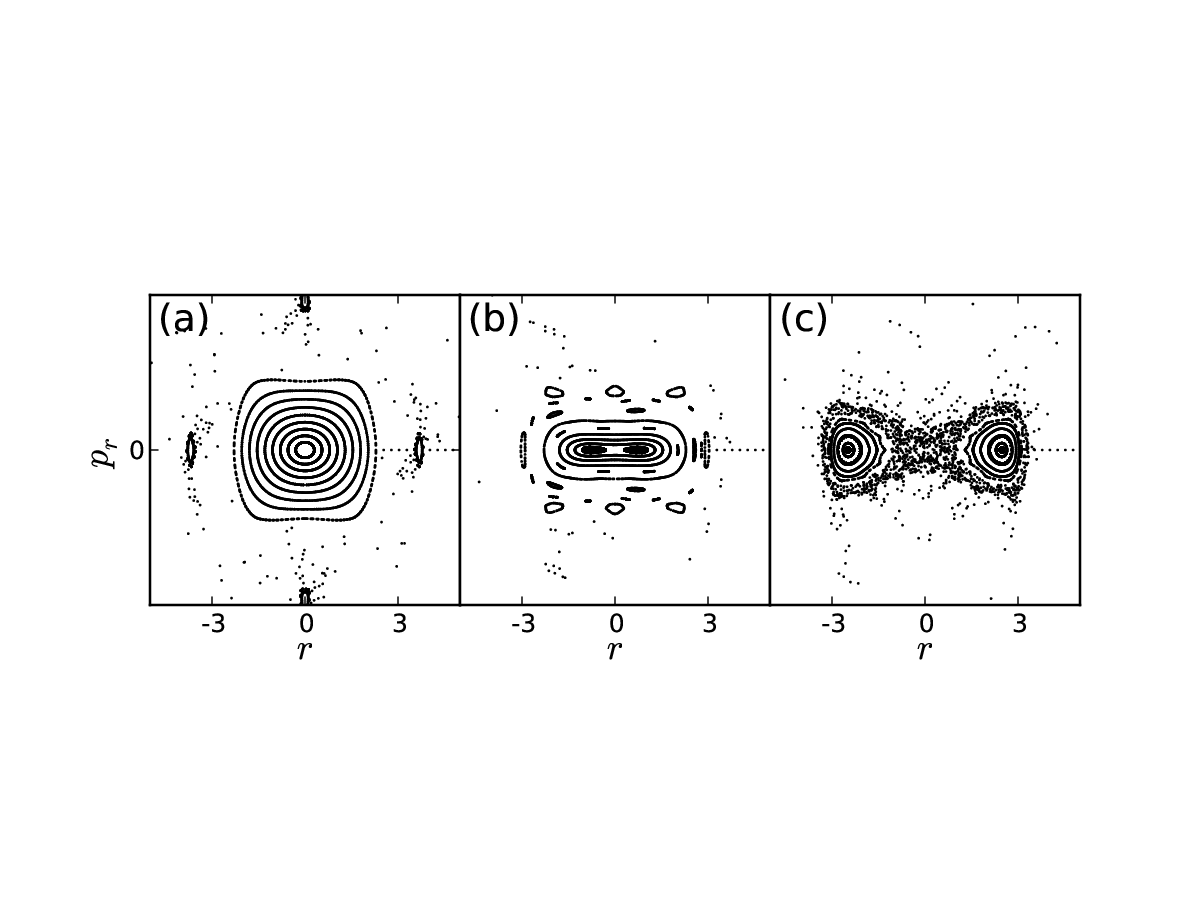}
\caption{\label{poincare}Three Poincaré sections corresponding respectively to points $A$, $B$ and $C$ on Fig.~\ref{stab}. The abscissa represents the radial position while the ordinate is the radial momentum of the particles (dimensionless). As we move from $A$ to $C$, we observe a bifurcation toward chaos by the period doubling mechanism.}
\end{figure}

The first major result of this paper is that configurations where trapping can be observed experimentally are mainly contained in the stability region defined by Floquet's theory. However, besides the inaccuracy near (4500, 5400) due to the 1\% error on the analytical potential \cite{VI2010}, we can see on Fig.~\ref{stab}, that some settings (e.g., in the region marked $C$) are theoretically predicted to be stable, and are not observed experimentally. This phenomenon can be explained by the non-linear couplings that appear when Eq.~\ref{quadpot} is not truncated and which are not taken into account in the previous linear approximation. We have numerically integrated the trajectories of ions taking into account the non-linear term in $r^2$ of equation \eqref{E:gp1}. The dynamic appears to be chaotic in the upper part of the right hand side stability region. To illustrate this, we have plotted on Fig.~\ref{poincare} three Poincaré sections corresponding to the three points denoted $A$, $B$ and $C$ on Fig.~\ref{stab}. The division in two islands separated by a stochastic region is the signature of a bifurcation toward chaos by a period doubling process. This is confirmed by the fact that the distance $d_r$ between the two islands grows as the square root of the parameter ($d_r=0.0054\times \sqrt{V_1-6192}$) \cite{HAK1991}. This explains that, theoretically, trapping is possible in all the areas predicted by the linear theory, even if the chaotic regime is more difficult to observe experimentally.

\begin{figure}[tbp]
\includegraphics[width=\columnwidth,trim =0cm 0.1cm 0cm 0cm,clip]{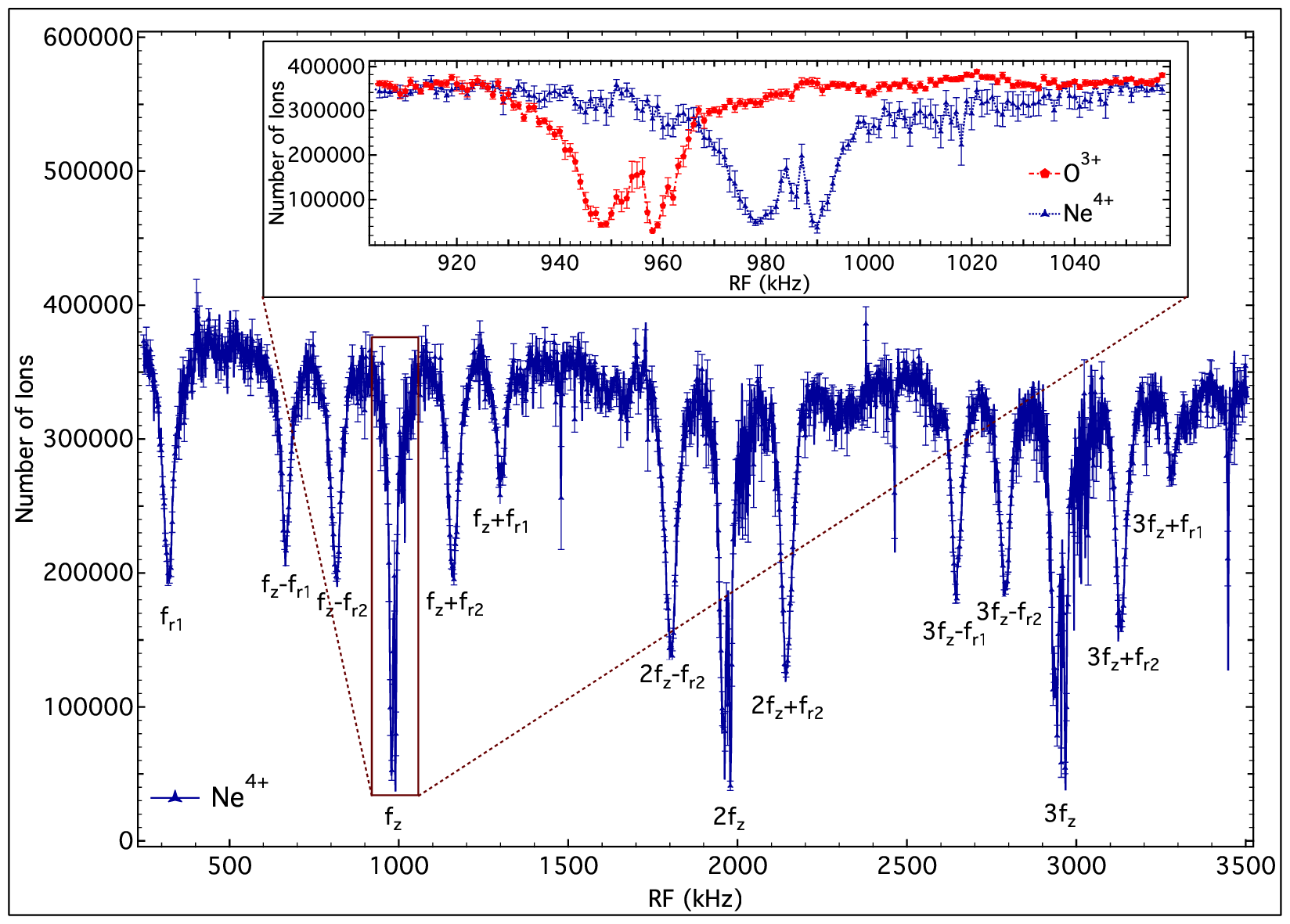}
\caption{\label{parametric} Experimental observation of parametric resonances when $f_{ex}=nf_z+mf_r$ with $n$ and $m$ integers for $Ne^{4+}$ ions at \unit{5.2}{keV/charge}. This is the perfect equivalent of what has been achieved with Paul traps \cite{CHAA1998}. As on Fig.~\protect\ref{stab}, each point represents the average of the ion signal on the faraday cup over several trapping cycles of \unit{500}{$\mu$s} for each frequency. The insert shows a magnification of the first main parametric resonance of the displayed frequency spectrum, along with the scan of the first resonance of $O^{3+}$ ions on the same scale.}
\end{figure}

We have discussed the stability with no RF excitation  ($V_{ex}=0$). We now study the EIBT behavior when varying the frequency of the sinusoidal excitation signal for $V_{ex}=\unit{10}{V}$ with a fixed set of electrode potentials ($V_1=$ \unit{6} {kV}, $V_2=$ \unit{5.825} {kV}, $V_3=$ \unit{4.125} {kV}, $V_4=$ \unit{1.65} {kV} and $V_z=$\unit{4.785} {kV}). When the excitation frequency is resonant with the ions motion frequency, ions gain energy at each oscillation and their trajectory becomes unstable as shown in \cite{CHA1998}. Fig.~\ref{parametric} illustrates the parametric resonances of Ne$^{4+}$ ions that can be observed when scanning the $V_{ex}$ excitation frequency. On the insert, next to the zoom of the first resonance of the Ne$^{4+}$ ions, the first resonance of O$^{3+}$ ions from a similar scan is shown. The shift in frequency is equivalent to the different charge/mass ratios, proving, for the first time, the ability to use EIBT as a mass spectrometer in this manner. Each peak is unfortunately divided in two due to a $\Delta x\simeq $\unit{1} {mm} shift of the excitating electrode from the center, resulting in a peak division with a $\Delta f\simeq \frac{2f^2\Delta x}{v}$ separation (where $v$ is the velocity of the ion in the field free region.)
This kind of resonance is not only useful to monitor the number of ions in the trap, but also leads to a new mass spectrometry method with interesting advantages.

The EIBT has already been used as TOF mass spectrometer \cite{sgp2002}. This technique can be compared to what is achieved with Penning traps, where ions oscillations are recorded to be analyzed by Fourier transform, giving a mass spectrum \cite{MGW09}. This is only possible if the potentials are chosen in order to be in the ``synchronization-mode'' \cite{psa2002}, where the ions stay bunched, resisting the coulomb repulsion. This counterintuitive phenomenon is explained in \cite{sgp2002} and requires a minimum number of trapped ions (few millions).

Synchronization restricts the use of the EIBT as a mass spectrometer in two ways. First, the unavoidable ion losses during trapping limit the time of synchronization and thus of observation, leading to a reduced upper bound of the mass spectrum's resolution. Second, as  mentioned in Ref.~\cite{sgp2002}, synchronization tends to aggregate in the same bunch species whose charge/mass ratio are close. This is known in mass spectrometry as ``peak-coalescence'' \cite{mahc1998} and limits the resolving power.

The results presented above suggest that an EIBT can be used for mass spectrometry in a similar way as quadrupole traps, where each ion species is successively ejected from the trap with a parametric excitation \cite{MGW09}. Usual methods (e.g., electrospray) could be used to  ionize samples and send the resulting ions  into  an EIBT. Since the oscillation frequency is proportional to the square root of the charge/mass ratio, sweeping the excitation frequency $f_{ex}$ provides a mass spectrum of the desired range. Besides the simplicity of the setup, this technique has two fundamental advantages. First, it is independent of synchronization and thus is limited neither by synchronization time nor by peak-coalescence. Second, the stability diagram shown on Fig.~\ref{stab} is computed with Eq. ~\eqref{hill}, whose only parameters are the ratios of the trapping potentials to the accelerating potential $V_i/E$. This means that all the charged particles injected in the trap with the given accelerating potential will be trapped independently of their mass and charge. This is very different from quadrupole traps where the stability of the trapping depends on the charge/mass ratio, limiting the range of the species that can be trapped and analyzed simultaneously. 

The resonances illustrated on Fig.~\ref{parametric} are broad and the resolution achieved ($M/\delta M\simeq100)$ cannot compete with advanced techniques. However, many parameters can be optimized and all the techniques dedicated to the quadrupole mass filters could be applied to enhance the resolution. For instance, we have noticed that the peaks are narrower when $V_z$ is tuned or when the radial extent of the injected beam is reduced by closing a set of upstream vertical slits. Moreover, the unlimited mass range could be profitable to chemists for analyzing molecules and in this case the achieved resolution is enough. 

The here described theoretical model can be applied directly to the ConeTrap \cite{SCJ2001} and electrostatic rings \cite{mol1997,bmbt2008} since the only requirement for the radial motion to be governed by a Hill's equation is to have ions moving periodically in an electrostatic field. The theoretical description and the analogy with the quadrupolar and other RF traps will well serve also the auto resonant trap (ART) community \cite{ERH2010}.

In conclusion, there are three major results in this work. First, the stability map, which is an experimental validation of the theoretical model based on Hill's equation. We think that we have thus provided a solid theoretical basis for future studies of this kind of traps. Second, we have shown that there is chaos in EIBT making it a new experimental tool to study non linear dynamics. Third, we have proven that parametric resonances enables to use the EIBT as a mass spectrometer and that its range in mass is not limited: ions with hugh difference in mass can be stored and analyzed simultaneously. Our experimental results suffer of a few artifacts, but we hope our approach will be confirmed by other users of such traps and will reveal itself as useful as the Mathieu equation for Paul trap users.

Laboratoire Kastler Brossel is ``UMR n$^{\circ}$ 8552'' (CNRS, ENS, UPMC).
The SIMPA ECRIS is supported by the infrastructure plan from UPMC. These experiments are supported by grant number \emph{ANR-06-BLAN-0233} of the ``Agence Nationale pour la Recherche (ANR)'' , and Helmholtz Alliance HA216/EMMI. We thank K.-D. Nguyen Thu-Lam for his help on the numerical calculations and J.P.~Okpisz, B. Delamour, P. Travers and J.M. Isac for technical support and the ASUR team from INSP, with whom we share the SIMPA ECRIS.

\bibliography{ref}

\end{document}